\documentclass[prb,preprintnumbers,amsmath,amssymb,floatfix]{revtex4}

\usepackage{graphicx}
\usepackage{subfigure}
\usepackage{dcolumn}

\begin{document}

\title{Two-dimensional function photonic crystals}
\author{Xiang-Yao Wu$^{a}$ \footnote{E-mail: wuxy2066@163.com},
 Ji Ma$^{a}$, Xiao-Jing Liu$^{a}$ and Yu Liang$^{a}$ }
 \affiliation{a. Institute of Physics, Jilin Normal
University, Siping 136000 China}

\begin{abstract}
In this paper, we have firstly proposed two-dimensional function
photonic crystals, which the dielectric constants of medium
columns are the functions of space coordinates $\vec{r}$, it is
different from the two-dimensional conventional photonic crystals
constituting by the medium columns of dielectric constants are
constants. We find the band gaps of two-dimensional function
photonic crystals are different from the two-dimensional
conventional photonic crystals, and when the functions form of
dielectric constants are different, the band gaps structure should
be changed, which can be designed into the appropriate band gaps
structures by the two-dimensional function photonic crystals.

\vskip 10pt


\end{abstract}

\vskip 10pt \maketitle {\bf 1. Introduction} \vskip 10pt

Photonic crystals (PCs) have generated a surge of interest in the
last decades because they offer the possibility to control the
propagation of light to an unprecedented level [1-4]. In its
simplest form, a photonic crystal is an engineered inhomogeneous
periodic structure made of two or more materials with very
different dielectric constants. PCs important characteristics are:
photon band gap, defect states, light localization and so on.
These characteristics make it able to control photons, so it may
be used to manufacture some high performance devices which have
completely new principles or can not be manufactured before, such
as high-efficiency semiconductor lasers, right emitting diodes,
wave guides, optical filters, high-Q resonators, antennas,
frequency-selective surface, optical wave guides and sharp bends
[5, 6], WDM-devices [7, 8], splitters and combiners [9, 10].
optical limiters and amplifiers [11, 12]. The research on photonic
crystals will promote its application and development on
integrated photoelectron devices and optical communication.

It is matter of general knowledge that by now the PCs can
constructed by varying photonic forbidden band (FB) of a structure
in one-, two- and three-dimensional photonic crystals [13-15]. The
optical and electronic properties of two- and three -dimensional
are intensively studied with the goal of achieving control of
electromagnetic propogation, and, especially, a complete photonic
band gap in all directions [16-18].

Owing to the PCs periodicity, the plane-wave expansion (PWE)
method is conventional for calculating PCs modes and photonic band
structures. Its essence is the Fourier expansion of
electromagnetic fields and material parameters, which is not only
a counterpart of the PWE method for electronic crystals, but it
also advantageously follows the classical coupled-wave theory
developed for diffraction gratings [19-21].

In Refs. [22-28], we have proposed one-dimensional function
photonic crystals, which is constituted by two media $A$ and $B$,
their refractive indexes are the functions of space position.
Unlike conventional photonic crystal (PCs), which is constituted
by the constant refractive index media $A$ and $B$. We have
studied the transmissivity and the electric field distribution
with and without defect layer. In this paper, we have firstly
proposed two-dimensional function photonic crystals, which the
dielectric constants of medium columns are the functions of space
coordinates $\vec{r}$, it is different from the two-dimensional
conventional photonic crystals constituting by the medium columns
of dielectric constants are constants. We find the band gaps of
two-dimensional function photonic crystals are different from the
two-dimensional conventional photonic crystals, and when the
functions form of dielectric constants are different, the band
gaps structure should be changed, which can be designed into the
appropriate band gaps structures by the two-dimensional function
photonic crystals.

\newpage

 {\bf 2. The plane-wave expansion method of two-dimensional photonic crystals} \vskip 8pt

When the medium constituting photonic crystals is passive, the
Maxwell equations are
\begin{eqnarray}
\nabla\cdot \vec{D}=0,
\end{eqnarray}
\begin{eqnarray}
\nabla\cdot \vec{B}=0,
\end{eqnarray}
\begin{eqnarray}
\nabla\times \vec{E}=-\mu_0\frac{\partial \vec{H}}{\partial t},
\end{eqnarray}
\begin{eqnarray}
\nabla\times \vec{H}=\varepsilon_0\varepsilon
(\vec{r})\frac{\partial \vec{E}}{\partial t},
\end{eqnarray}
where $\varepsilon (\vec{r})$ is the position-dependent dielectric
constant.

For the monochromatic plane electromagnetic wave, the electric
field and magnetic field intensity are
\begin{eqnarray}
\vec{H}(\vec{r},t)=\vec{H}(\vec{r})e^{-i\omega t},
\end{eqnarray}
\begin{eqnarray}
\vec{E}(\vec{r},t)=\vec{E}(\vec{r})e^{-i\omega t}.
\end{eqnarray}
For the two-dimensional photonic crystals, the medium columns of
dielectric constant $\varepsilon_a (\vec{r})$ are located in the
background of dielectric constant $\varepsilon_b$, and the medium
columns are parallel the z-axis, the cross section form
two-dimensional grids. In the following, we should give the eigenvalue equations of $TM$ and $TE$ wave.\\

(1) For the $TM$ wave, the electric field and magnetic field
intensity are
\begin{eqnarray}
\vec{H}(\vec{r},t)=(0,0,H_z(\vec{r}))e^{-i\omega t},
\end{eqnarray}
\begin{eqnarray}
\vec{E}(\vec{r},t)=(E_x(\vec{r}),E_y(\vec{r}),0)e^{-i\omega t},
\end{eqnarray}
where $\vec{r}=x\vec{i}+y\vec{j}$.

Substituting Eqs. (7) and (8) into (3) and (4), we obtain
\begin{eqnarray}
\frac{\partial E_y}{\partial x}-\frac{\partial E_x}{\partial
y}=i\omega \mu_0H_z,
\end{eqnarray}
\begin{eqnarray}
\frac{\partial H_z}{\partial x}=i\omega \varepsilon_0
\varepsilon(\vec{r}) E_y,
\end{eqnarray}
\begin{eqnarray}
\frac{\partial H_z}{\partial y}=-i\omega \varepsilon_0
\varepsilon(\vec{r}) E_x.
\end{eqnarray}
In Eqs. (9)-(11), we take out $E_x$ and $E_y$, and obtain the
$H_z$ equation
\begin{eqnarray}
\frac{\partial}{\partial x}[\frac{1}{\varepsilon
(\vec{r})}\frac{\partial H_z(\vec{r})}{\partial
x}]+\frac{\partial}{\partial y}[\frac{1}{\varepsilon
(\vec{r})}\frac{\partial H_z(\vec{r})}{\partial
y}]=-\frac{\omega^2}{c^2}H_z(\vec{r}).
\end{eqnarray}
In the periodic dielectric of two-dimensional photonic crystals,
 the dielectric constant $\varepsilon (\vec{r})$ satisfy
\begin{eqnarray}
\varepsilon (\vec{r}+\vec{R})=\varepsilon (\vec{r}),
\end{eqnarray}
where $\vec{R}=l_1\vec{a}_1+l_2\vec{a}_2$ is the lattice vector,
the vectors $\vec{a}_1$ and $\vec{a}_2$ are the basis vector of
cell, and $l_1$ and $l_2$ are arbitrary integers. We introduce the
cell basis vectors $\vec{b}_1$ and $\vec{b}_2$ of reciprocal
space, and reciprocal lattice vector $\vec{G}$, they are
\begin{eqnarray}
\vec{a}_i\cdot\vec{b}_j=2\pi\delta_{ij}\hspace{0.3in}(i,j=1,2),
\end{eqnarray}
\begin{eqnarray}
\vec{G}=m_1\vec{b}_1+m_2\vec{b}_2\hspace{0.3in}(m_1,m_2=0,\pm1,\pm2,\cdot\cdot\cdot).
\end{eqnarray}
In the periodic medium, magnetic field intensity $H_z(\vec{r})$ is
periodic distribution and satisfy the Block law
\begin{eqnarray}
\left \{ \begin{array}{ll}
H_z(\vec{r})=H_k(\vec{r})e^{i\vec{k}\cdot\vec{r}}\\
H_k(\vec{r}+\vec{R})=H_k(\vec{r}),
\end{array}
\right.
\end{eqnarray}
where $k$ is the Block wave vector.

We can spread $H_z(\vec{r})$ and $\frac{1}{\varepsilon(\vec{r})}$
as the Fourier series in the reciprocal space, they are
\begin{eqnarray}
H_k(\vec{r})=\sum_{\vec{G}} H_k(\vec{G})e^{i\vec{G}\cdot\vec{r}},
\end{eqnarray}
\begin{eqnarray}
\frac{1}{\varepsilon (\vec{r})}=\sum_{\vec{G'}}
{\varepsilon}^{-1}(\vec{G'})e^{i\vec{G'}\cdot\vec{r}},
\end{eqnarray}
substituting Eq. (17) into (16), we have
\begin{eqnarray}
H_z(\vec{r})=\sum_{\vec{G}}
H_k(\vec{G})e^{i(\vec{k}+\vec{G})\cdot\vec{r}},
\end{eqnarray}
substituting Eqs. (18) and (19) into (12), we get
\begin{eqnarray}
&&\frac{\partial}{\partial
x}[\sum_{\vec{G''}}\varepsilon^{-1}(\vec{G''})e^{i\vec{G''}\cdot\vec{r}}
\sum_{\vec{G'}}i(k_x+G_x^{'})H_{\vec{k}}(\vec{G'})e^{i(\vec{k}+\vec{G'})\cdot\vec{r}}]
\nonumber\\&&+\frac{\partial}{\partial
y}[\sum_{\vec{G''}}\varepsilon^{-1}(\vec{G''})e^{i\vec{G''}\cdot\vec{r}}
\sum_{\vec{G'}}i(k_y+G_y^{'})H_{\vec{k}}(\vec{G'})e^{i(\vec{k}+\vec{G'})\cdot\vec{r}}]\nonumber
\\&&=-\frac{\omega^2}{c^2}\sum_{\vec{G}}H_{\vec{k}}(\vec{G})e^{i(\vec{k}+\vec{G})\cdot\vec{r}},
\end{eqnarray}
i.e.,
\begin{eqnarray}
&&\sum_{\vec{G'},\vec{G''}}(k_x+G_x^{'})(k_x+G_x^{'}+G_x^{''})\varepsilon^{-1}(\vec{G''})
H_{\vec{k}}(\vec{G'})e^{i(\vec{k}+\vec{G'}+\vec{G''})\cdot\vec{r}}
\nonumber\\&&+\sum_{\vec{G'},\vec{G''}}(k_y+G_y^{'})(k_y+G_y^{'}+G_y^{''})\varepsilon^{-1}(\vec{G''})
H_{\vec{k}}(\vec{G'})e^{i(\vec{k}+\vec{G'}+\vec{G''})\cdot\vec{r}}\nonumber
\\&&=\frac{\omega^2}{c^2}\sum_{\vec{G}}H_{\vec{k}}(\vec{G})e^{i(\vec{k}+\vec{G})\cdot\vec{r}},
\end{eqnarray}
taking $\vec{G'}+\vec{G''}=\vec{G}$, the Eq. (21) can be written
as
\begin{eqnarray}
&&\sum_{\vec{G},\vec{G'}}(k_x+G_x^{'})(k_x+G_x)\varepsilon^{-1}(\vec{G}-\vec{G'})
H_{\vec{k}}(\vec{G'})e^{i(\vec{k}+\vec{G})\cdot\vec{r}}
\nonumber\\&&+\sum_{\vec{G},\vec{G'}}(k_y+G_y^{'})(k_y+G_y)\varepsilon^{-1}(\vec{G}-\vec{G'})
H_{\vec{k}}(\vec{G'})e^{i(\vec{k}+\vec{G})\cdot\vec{r}}\nonumber
\\&&=\frac{\omega^2}{c^2}\sum_{\vec{G}}H_{\vec{k}}(\vec{G})e^{i(\vec{k}+\vec{G})\cdot\vec{r}},
\end{eqnarray}
since $e^{i(\vec{k}+\vec{G})\cdot\vec{r}}$ is independent for
every $\vec{G}$, the both sides coefficients of Eq. (22) are equal
for different $\vec{G}$, there is
\begin{eqnarray}
\sum_{\vec{G'}}((k_x+G_x^{'})(k_x+G_x)+(k_y+G_y^{'})(k_y+G_y))\varepsilon^{-1}(\vec{G}-\vec{G'})
H_{\vec{k}}(\vec{G'}) =\frac{\omega^2}{c^2}H_{\vec{k}}(\vec{G}),
\end{eqnarray}
i.e.,
\begin{eqnarray}
\sum_{\vec{G'}}(\vec{k}+\vec{G}^{'})\cdot(\vec{k}+\vec{G})\varepsilon^{-1}(\vec{G}-\vec{G'})
H_{\vec{k}}(\vec{G'}) =\frac{\omega^2}{c^2}H_{\vec{k}}(\vec{G}).
\end{eqnarray}
The Eq. (24) is the eigenvalue equation of $TM$ wave.

(1) For the $TE$ wave, the electric field and magnetic field
intensity are
\begin{eqnarray}
\vec{E}(\vec{r},t)=(0,0,E_z(\vec{r}))e^{-i\omega t},
\end{eqnarray}
\begin{eqnarray}
\vec{H}(\vec{r},t)=(H_x(\vec{r}),H_y(\vec{r}),0)e^{-i\omega t},
\end{eqnarray}
substituting Eqs. (25) and (26) into (3) and (4), we obtain
\begin{eqnarray}
\frac{\partial H_y}{\partial x}-\frac{\partial H_x}{\partial
y}=-i\omega \varepsilon_0 \varepsilon(\vec{r}) E_z,
\end{eqnarray}
\begin{eqnarray}
\frac{\partial H_z}{\partial x}=-i\omega \mu_0 H_y,
\end{eqnarray}
\begin{eqnarray}
\frac{\partial E_z}{\partial y}=-i\omega \mu_0 H_x.
\end{eqnarray}
In Eqs. (27)-(29), we take out $H_x$ and $H_y$, and obtain the
$E_z$ equation
\begin{eqnarray}
\frac{1}{\varepsilon (\vec{r})}[\frac{\partial^2}{\partial
x^2}+\frac{\partial^2}{\partial
y^2}]E_z(\vec{r})=-\frac{\omega^2}{c^2}H_z(\vec{r}),
\end{eqnarray}
similarly Eq. (19), $E_z(\vec{r})$ can be written as
\begin{eqnarray}
E_z(\vec{r})=\sum_{\vec{G}}
E_k(\vec{G})e^{i(\vec{K}+\vec{G})\cdot\vec{r}}.
\end{eqnarray}
By the the same steps of Eqs. (18)-(22), we can obtain the
eigenvalue equation of $TE$ wave
\begin{eqnarray}
\sum_{\vec{G'}}|\vec{k}+\vec{G}^{'}||\vec{k}+\vec{G}|\varepsilon^{-1}(\vec{G}-\vec{G'})
E_{\vec{k}}(\vec{G'}) =\frac{\omega^2}{c^2}E_{\vec{k}}(\vec{G}).
\end{eqnarray}
The Eqs. (24) and (32) are suited to both two-dimensional
conventional photonic crystals and two-dimensional function
photonic crystals

\vskip 8pt
 {\bf 3. The Fourier transform of dielectric constant for two-dimensional function photonic crystals} \vskip 8pt
For the two-dimensional function photonic crystals, the medium
column dielectric constants are the function of  space coordinates
$\vec{r}$, it is different from the two-dimensional conventional
photonic crystals, which dielectric constants are constant. The
dielectric constant of cylindrical medium column can be written as
\begin{eqnarray}
\varepsilon(\vec{r})=\left \{\begin{array}{ll}
\varepsilon_{a}(\vec{r})\hspace{0.2in} r\leq r_{a}
\\ \varepsilon_{b}\hspace{0.2in} r>r_{a}
    \end{array}
   \right.,
\end{eqnarray}
or
\begin{eqnarray}
\frac{1}{\varepsilon(\vec{r})}=\left
\{\begin{array}{ll}\frac{1}{\varepsilon_{a}(\vec{r})}\hspace{0.2in}
r\leq r_{a}
\\\frac{1}{\varepsilon_{b}}\hspace{0.2in} r>r_{a}
    \end{array}
   \right.,
\end{eqnarray}
Eq. (34) can be written as
\begin{eqnarray}
\frac{1}{\varepsilon(\vec{r})}=\frac{1}{\varepsilon_{b}}+(\frac{1}{\varepsilon_{a}(\vec{r})}-\frac{1}{\varepsilon_{b}})s(r)
\end{eqnarray}
where
\begin{eqnarray}
s(r)=\left \{\begin{array}{ll}1\hspace{0.2in} r\leq r_{a}
\\0\hspace{0.2in} r>r_{a}
    \end{array}
   \right..
\end{eqnarray}
The Fourier inverse transform of $\frac{1}{\varepsilon(\vec{r})}$
is
\begin{eqnarray}
\varepsilon^{-1}(\vec{G})=\frac{1}{V_{0}}\int_{V_{0}}d\vec{r}\frac{1}{\varepsilon(\vec{r})}e^{-i\vec{G}\cdot\vec{r}},
\end{eqnarray}
in the two-dimensional reciprocal space, it is
\begin{eqnarray}
\varepsilon^{-1}(\vec{G}_{||})=\frac{1}{V_{0}^{(2)}}\int_{V_{0}^{(2)}}d\vec{r}_{||}\frac{1}{\varepsilon(\vec{r}_{||})}e^{-i\vec{G}_{||}\cdot\vec{r}_{||}},
\end{eqnarray}

where $\vec{G}_{||}=m_1\vec{b}_1+m_2\vec{b}_2$,
$\vec{r}_{||}=x\vec{i}+y\vec{j}$, $V_{0}^{(2)}$ represents the
unit cell area in the two dimensional lattice space.\\

Substituting Eq. (35) into (38), there is
\begin{eqnarray}
\varepsilon^{-1}(\vec{G}_{||})&&=\frac{1}{V_{0}^{(2)}}\int_{V_{0}^{(2)}}d\vec{r}_{||}[\frac{1}{\varepsilon_{b}}+(\frac{1}{\varepsilon_{a}}-\frac{1}{\varepsilon
_{b}})s(\vec{r}_{||})]e^{-i\vec{G}_{||}\cdot\vec{r}_{||}}\nonumber
\\&&=\frac{1}{\varepsilon _{b}}\frac{1}{V_{0}^{(2)}}
V_{0}^{(2)}\delta_{\vec{G}_{||,0
}}+\frac{1}{V_{0}^{(2)}}\int_{V_{0}^{(2)}}d\vec{r}_{||}(\frac{1}{\varepsilon_{a}}-\frac{1}{\varepsilon
_{b}})s(\vec{r}_{||})e^{-i\vec{G}_{||}\cdot\vec{r}_{||}}\nonumber
\\&&=\frac{1}{\varepsilon _{b}}\delta_{m,0}
\delta_{n,0}+\frac{1}{V_{0}^{(2)}}\int_{V_{0}^{(2)}}d\vec{r}_{||}(\frac{1}{\varepsilon_{a}}-\frac{1}{\varepsilon
_{b}})s(\vec{r}_{||})e^{-i\vec{G}_{||}\cdot\vec{r}_{||}}\nonumber
\\&&=\frac{1}{\varepsilon _{b}}\delta_{m,0}
\delta_{n,0}+I,
\end{eqnarray}
where

\begin{eqnarray}
I&&=\frac{1}{V_{0}^{(2)}}\int_{V_{0}^{(2)}}d\vec{r}_{||}(\frac{1}{\varepsilon_{a}}-\frac{1}{\varepsilon
_{b}})s(\vec{r}_{||})e^{-i\vec{G}_{||}\cdot\vec{r}_{||}}\nonumber\\&&=\frac{1}{V_{0}^{(2)}}\int_{V_{0}^{(2)}}d\vec{r}_{||}
\frac{1}{\varepsilon_{a}}s(\vec{r}_{||})e^{-i\vec{G}_{||}\cdot\vec{r}_{||}}-\frac{1}{V_{0}^{(2)}}\int_{V_{0}^{(2)}}d\vec{r}_{||}
\frac{1}{\varepsilon_{b}}s(\vec{r}_{||})e^{-i\vec{G}_{||}\cdot\vec{r}_{||}}\nonumber\\&&=I_{1}-I_{2}
\end{eqnarray}

where

\begin{eqnarray}
I_{2}&&=\frac{1}{V_{0}^{(2)}}\int_{V_{0}^{(2)}}d\vec{r}_{||}
\frac{1}{\varepsilon_{b}}s(\vec{r}_{||})e^{-i\vec{G}_{||}\cdot\vec{r}_{||}}
\nonumber\\&&=\frac{1}{\varepsilon_{b}}\frac{1}{V_{0}^{(2)}}\int_{V_{0}^{(2)}}d\vec{r}_{||}
s(\vec{r}_{||})e^{-i\vec{G}_{||}\cdot\vec{r}_{||}}
\nonumber\\&&=\frac{1}{\varepsilon_{b}}\frac{1}{V_{0}^{(2)}}\int^{r_{a}}_{0}r
dr\int^{2\pi}_{0}d\theta e^{-iG_{||}\cdot r\cdot
cos\theta}\nonumber\\&&=\frac{1}{\varepsilon_{b}}\frac{1}{V_{0}^{(2)}}\int^{r_{a}}_{0}r
dr\int^{2\pi}_{0} d\theta e^{iG_{||}\cdot r \cdot
sin(\theta-\frac{\pi}{2})}
\end{eqnarray}
where $|\vec{r}_{||}|={r}_{||}=r$, $|\vec{G}_{||}|={G}_{||}$,
$d\vec{r}_{||}=ds=rdrd\theta$, and $\theta$ is the included angle
of $\vec{r}_{||}$ and $\vec{G}_{||}$.

By the formulas
\begin{eqnarray}
e^{i\omega\cdot
sin\theta}=\sum\limits^{\infty}_{l=-\infty}J_{l}(\omega)e^{il\theta},
\end{eqnarray}
\begin{eqnarray}
\int^{2\pi}_{0}d \theta e^{il(\theta-\frac{\pi}{2})}=\left
\{\begin{array}{ll}0\hspace{0.28in} (l\neq0)
\\2\pi\hspace{0.2in} (l=0)
    \end{array}
   \right.,
\end{eqnarray}

and
\begin{eqnarray}
\int x^{m}J_{m-1}(x)dx=x^{m}J_{m}(x)+c,
\end{eqnarray}

\begin{figure}[tbp]
\includegraphics[width=16cm, height=8cm]{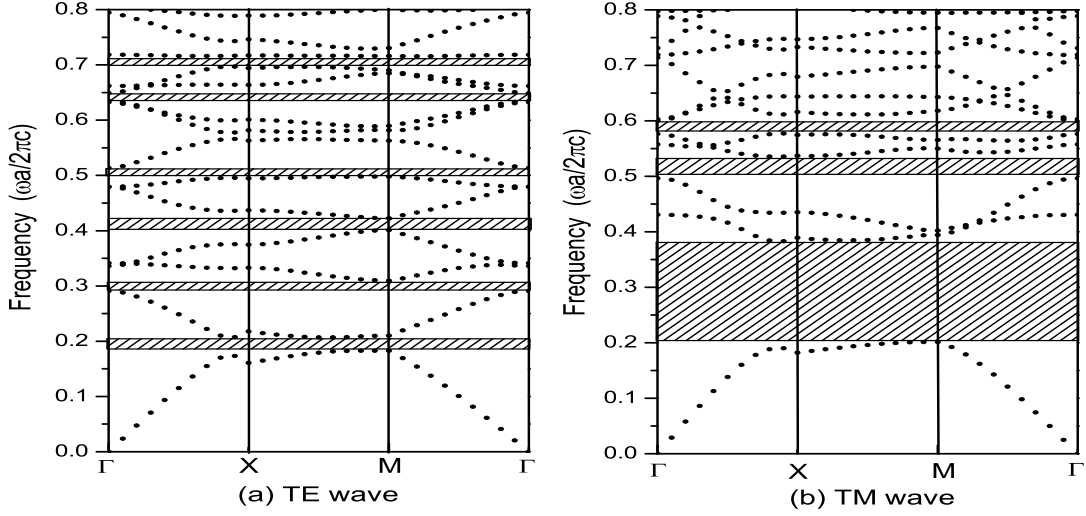}
\caption{The band gaps structure of two-dimensional conventional
photonic crystals for the triangle lattice, $\varepsilon_b=1$,
$\varepsilon_a=11.96$, medium column radius $r_a=0.4a$. (a) $TE$
wave, (b) $TM$ wave.}
\end{figure}
\begin{figure}[tbp]
\includegraphics[width=16cm, height=8cm]{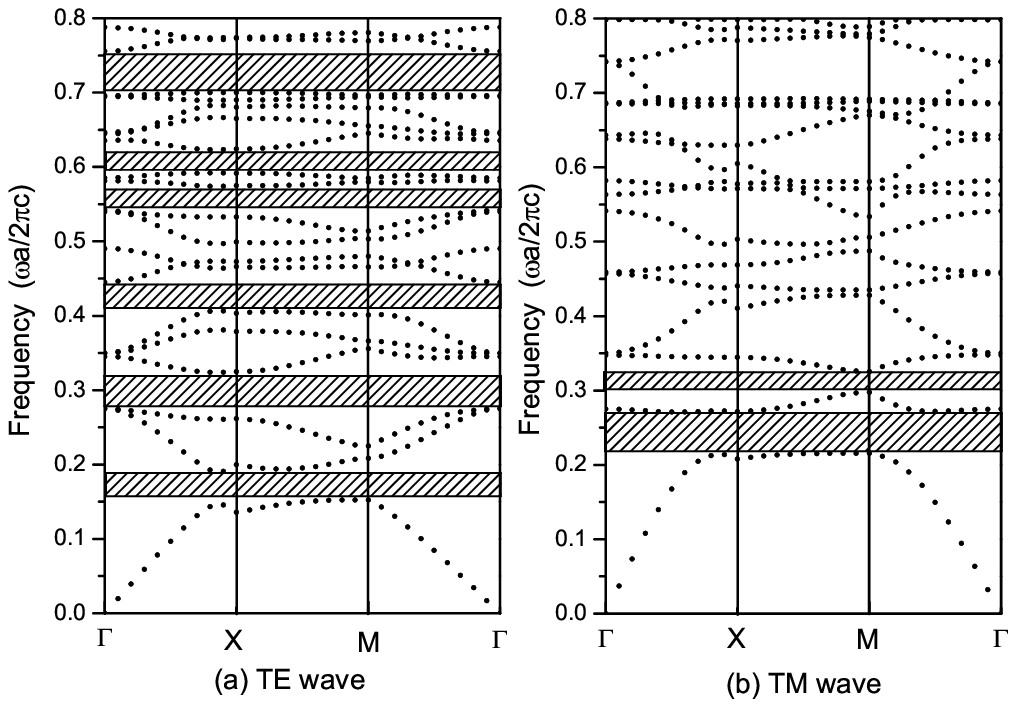}
\caption{The band gaps structure of two-dimensional function
photonic crystals for the triangle lattice, $\varepsilon_b=1$,
$\varepsilon_a=k\cdot r+11.96$, function coefficient $k=3\cdot
10^6$, medium column radius $r_a=0.4a$. (a) $TE$ wave, (b) $TM$
wave.}
\end{figure}
\begin{figure}[tbp]
\includegraphics[width=16cm, height=8cm]{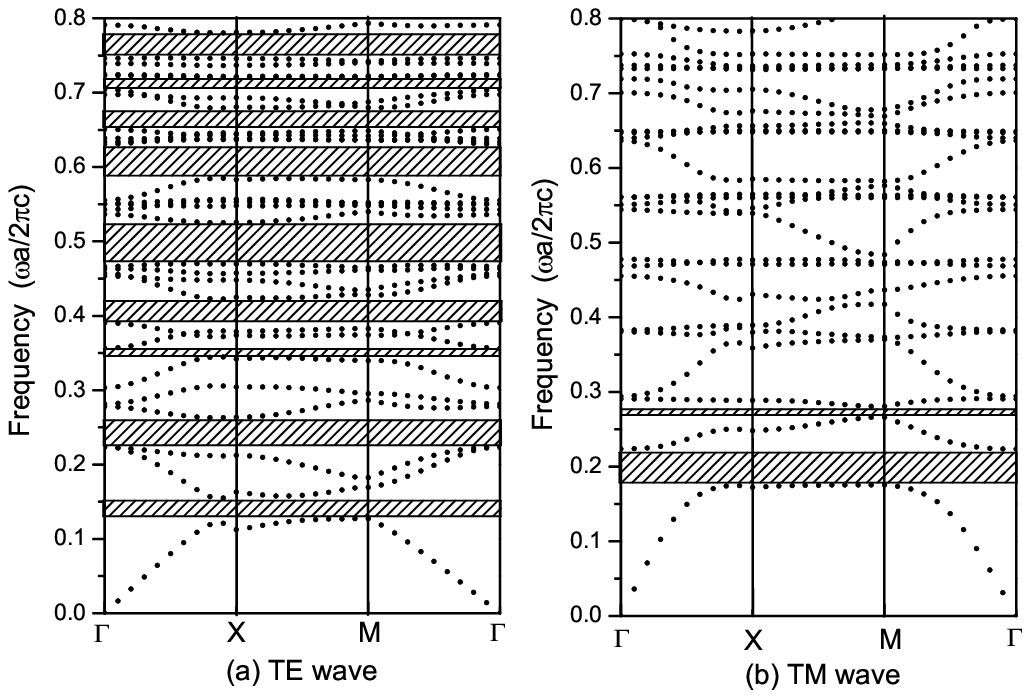}
\caption{The band gaps structure of two-dimensional function
photonic crystals for the triangle lattice, $\varepsilon_b=1$,
$\varepsilon_a=k\cdot r+11.96$, function coefficient $k=45\cdot
10^6$, medium column radius $r_a=0.4a$. (a) $TE$ wave, (b) $TM$
wave.}
\end{figure}
\begin{figure}[tbp]
\includegraphics[width=16cm, height=8cm]{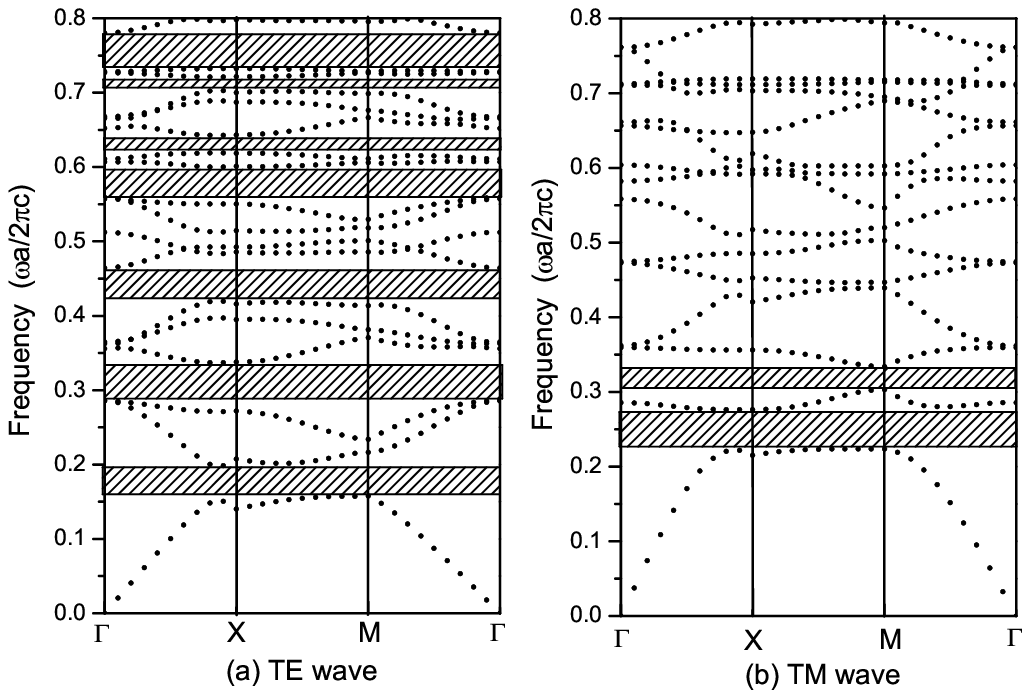}
\caption{The band gaps structure of two-dimensional function
photonic crystals for the triangle lattice, $\varepsilon_b=1$,
$\varepsilon_a=k\cdot r+11.96$, function coefficient $k=-3\cdot
10^6$, medium column radius $r_a=0.4a$. (a) $TE$ wave, (b) $TM$
wave.}
\end{figure}
\begin{figure}[tbp]
\includegraphics[width=16cm, height=8cm]{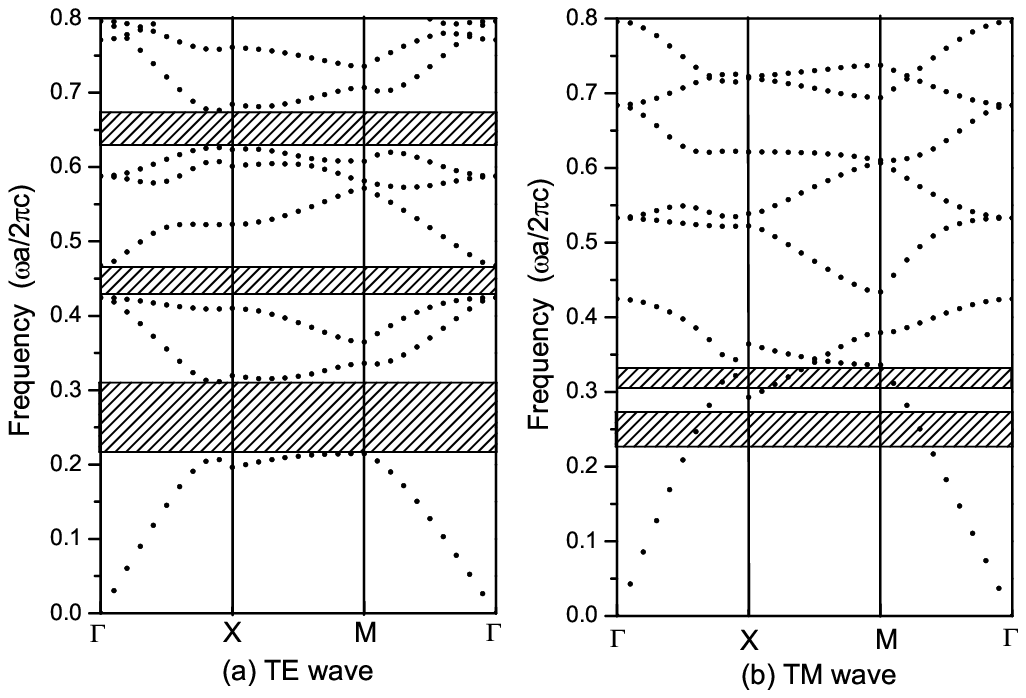}
\caption{The band gaps structure of two-dimensional function
photonic crystals for the triangle lattice, $\varepsilon_b=1$,
$\varepsilon_a=k\cdot r+11.96$, function coefficient $k=-45\cdot
10^6$, medium column radius $r_a=0.4a$. (a) $TE$ wave, (b) $TM$
wave.}
\end{figure}
we have
\begin{eqnarray}
I_{2}&&=\frac{1}{\varepsilon_{b}}\frac{1}{V_{0}^{(2)}}\int^{r_{a}}_{0}r
dr \sum\limits^{\infty}_{l=-\infty}J_{l}(G_{||}\cdot
r)\int^{2\pi}_{0}d\theta e^{il(\theta-\frac{\pi}{2})}
\nonumber\\&&=\frac{1}{\varepsilon_{b}}\frac{1}{V_{0}^{(2)}}\frac{2\pi
r_{a}}{G_{||}}\cdot J_{1}(G_{||}\cdot
r_{a}),\hspace{0.4in}(\vec{G}_{||}\neq 0)
\end{eqnarray}

when $G_{||}\rightarrow 0$ ($m\rightarrow 0$, $n\rightarrow 0$),
we have
\begin{eqnarray}
I_{2}(m=0,n=0)&&=\lim_{G_{||}\rightarrow
0}\frac{1}{\varepsilon_{b}}\frac{1}{V_{0}^{(2)}} 2\pi
r_{a}\frac{J_{1}(G_{||}\cdot
r_{a})}{G_{||}}\nonumber\\&&=\frac{1}{\varepsilon_{b}}\frac{1}{V_{0}^{(2)}}
2\pi r_{a}\lim_{G_{||}\rightarrow0}\frac{(J_{1}(G_{||}\cdot
r_{a}))^{\prime}}{(G_{||})^{\prime}}\nonumber\\&&=\frac{1}{\varepsilon_{b}}\frac{1}{V_{0}^{(2)}}
2\pi r_{a}\lim_{G_{||}\rightarrow 0}{J_{1}^{\prime}(G_{||}\cdot
r_{a})\cdot r_{a}}\nonumber\\&&=\frac{1}{\varepsilon_{b}}\frac{\pi
r_{a}^{2}}{V_{0}^{(2)}}\nonumber\\&&=\frac{f}{\varepsilon_{b}},\hspace{0.4in}(\vec{G}_{||}=
0)
\end{eqnarray}
where $J_1^{'}(0)=\frac{1}{2}$ and $f=\frac{\pi
r_{a}^{2}}{V_{0}^{(2)}}$ is filling ratio.

Where
\begin{eqnarray}
I_{1}&&=\frac{1}{V_{0}^{(2)}}\int_{V_{0}^{(2)}}d\vec{r}_{||}
\frac{1}{\varepsilon_{a}(r,\theta)}s(\vec{r}_{||})e^{-i\vec{G}_{||}\cdot\vec{r}_{||}}
\nonumber
\\&&=\frac{1}{V_{0}^{(2)}}\int^{r_{a}}_{0}r dr
\int^{2\pi}_{0}\frac{1}{\varepsilon_{a}(r,\theta)}e^{iG_{||}\cdot
r_{||}\cdot\sin (\theta-\frac{\pi}{2})}d\theta \nonumber
\\&&=\frac{1}{V_{0}^{(2)}}\int^{r_{a}}_{0}rdr\int^{2\pi}_{0}
d\theta
\frac{1}{\varepsilon_{a}(r,\theta)}\sum\limits^{\infty}_{l=-\infty}J_{l}(G_{||}\cdot
r)e^{il(\theta-\frac{\pi}{2})}\nonumber
\\&&=\frac{1}{V_{0}^{(2)}}\int^{r_{a}}_{0}r\frac{1}{\varepsilon_{a}(r)}\sum\limits^{\infty}_{l=-\infty}J_{l}(G_{||}\cdot
r)dr\int^{2\pi}_{0} d\theta e^{il(\theta-\frac{\pi}{2})} \nonumber
\\&&=\frac{2\pi}{V_{0}^{(2)}}\int^{r_{a}}_{0}
r\frac{1}{\varepsilon_{a}(r)}J_{0}(G_{||}\cdot
r)dr.\hspace{0.4in}(\vec{G}_{||}\neq 0)
\end{eqnarray}
In Eq. (47), we consider
$\varepsilon_{a}(r,\theta)=\varepsilon_{a}(r)$.

When $G_{||}=0$, as $J_0(0)=1$, we have
\begin{eqnarray}
I_{1}=\frac{2\pi}{V_{0}^{(2)}}\int^{r_{a}}_{0}
r\frac{1}{\varepsilon_{a}(r)}dr,\hspace{0.4in}(\vec{G}_{||}=0),
\end{eqnarray}
substituting $I_1$, $I_2$ and $I$ into Eq. (39), we obtain
\begin{eqnarray}
\varepsilon^{-1}(\vec{G}_{||})=\left
\{\begin{array}{ll}\frac{1}{\varepsilon_{b}}(1-f)+\frac{2f}{r_{a}^{2}}\int^{r_{a}}_{0}r\frac{1}{\varepsilon_{a}(r)}
dr\hspace{1.1in}(\vec{G}_{||}=0)
\\\frac{2f}{r_{a}^{2}}\int^{r_{a}}_{0}r\frac{1}{\varepsilon_{a}(r)}
J_{0}(G_{||}\cdot
r)dr-\frac{2f}{\varepsilon_{b}}\frac{J_{1}(G_{||}\cdot
r_{a})}{G_{||}\cdot r_{a}}\hspace{0.4in}(\vec{G}_{||}\neq 0)
    \end{array}
   \right.,
\end{eqnarray}
when $\varepsilon_{a}(r)=\varepsilon_{a}$, the $\varepsilon_{a}$
is a constant, the Eq. (49) becomes
\begin{eqnarray}
\varepsilon^{-1}(\vec{G}_{||})=\left
\{\begin{array}{ll}\frac{1}{\varepsilon_{b}}+(\frac{1}{\varepsilon_{a}}-\frac{1}{\varepsilon_{b}})f\hspace{1.1in}(\vec{G}_{||}=0)
\\2f(\frac{1}{\varepsilon_{a}}-\frac{1}{\varepsilon_{b}})\frac{J_1(G_{||}\cdot
r_{a})}{G_{||}\cdot r_{a}}\hspace{0.8in}(\vec{G}_{||}\neq 0)
    \end{array}
   \right..
\end{eqnarray}
The Eq. (50) is the dielectric constant Fourier transform of
two-dimensional conventional photonic crystals. So, the
two-dimensional conventional photonic crystals is the special case
of two-dimensional function photonic crystals.

\vskip 8pt {\bf 4. Numerical result} \vskip 8pt

In this section, we report our numerical results of band
structures for the two-dimensional function photonic crystals. In
order to compare the band gaps structures of the two-dimensional
conventional and function photonic crystals, we firstly calculate
the band gaps structures of the two-dimensional conventional
photonic crystals. The structure is the triangle lattice, and the
cylindrical medium column are located in the air, its dielectric
constant $\varepsilon_{a}=11.96$ and medium column radius
$r_a=0.4a$, where $a=10^{-6} m$ is lattice constant. The band gaps
structures of $TE$ and $TM$ wave are shown in FIG. 1. FIG. 1 (a)
and (b) are the band gaps structures of $TE$ and $TM$ wave,
respectively. In the frequency range of $0-0.8$ (in unit of
$a/2\pi c$), there are six narrow band gaps for the $TE$ wave, and
there are three band gaps, including a width band gap in the
frequency range of $0.2-0.38$ for the $TM$ wave.

In the following, we should calculate the band gaps structures of
the two-dimensional function photonic crystals. The structure is
the triangle lattice, and the cylindrical medium column are
located in the air, its dielectric constant is the function of
space coordinate, it is $\varepsilon_{a}(\vec{r})=kr+11.96 (0\leq
r\leq r_a)$ and medium column radius $r_a=0.4a$, the coefficient
$k$ is called function coefficient, when $k=0$, it is conventional
photonic crystals, when $k\neq 0$, it is function photonic
crystals. In FIG. 2 (a) and (b), we give the band gaps structures
of $TE$ and $TM$ wave when $k=3\cdot10^6$. In FIG. 2 (a), in the
frequency range of $0-0.8$, there are six band gaps for the $TE$
wave, which are wider than the FIG. 1 (a) band gaps (conventional
photonic crystals). In FIG. 2 (b), there are two band gaps for the
$TM$ wave, and they are narrower than the FIG. 1 (b) band gaps
(conventional photonic crystals). Near the $0.3$ frequency, the
band gaps of $TE$ and $TM$ wave appear overlap. In the other
frequency interval, the band gaps of $TE$ and $TM$ wave without
overlap, which can be designed the polarization selection device.
In FIG. 3 (a) and (b), we give the band gaps structures of $TE$
and $TM$ wave when the function coefficient $k=45\cdot10^6$. In
the frequency range of $0-0.8$, there are nine band gaps for the
$TE$ wave, and two band gaps for the $TM$ wave. Comparing with
FIG. 2, we can find the band gaps numbers of $TE$ wave increase
and red shift, and $TM$ wave band gaps become narrow and red shift
with the function coefficient $k$ increasing. Comparing with the
conventional photonic crystals FIG. 1, we can find the band gaps
numbers of $TE$ wave increase, become wider and red shift, and the
$TM$ wave band gaps numbers decrease, become narrow and red shift.
In FIG. 4 (a) and (b), we give the band gaps structures of $TE$
and $TM$ wave when the function coefficient $k=-3\cdot10^6$. In
the frequency range of $0-0.8$, there are seven band gaps for the
$TE$ wave, and two band gaps for the $TM$ wave. Comparing with the
conventional photonic crystals FIG. 1, we can find the band gaps
numbers of $TE$ wave increase and become wider, and the $TM$ wave
band gaps numbers decrease and become narrow. In FIG. 5 (a) and
(b), we give the band gaps structures of $TE$ and $TM$ wave when
the function coefficient $k=-45\cdot10^6$. In the frequency range
of $0-0.8$, there are three band gaps for the $TE$ wave, and one
band gap for the $TM$ wave. Comparing with FIG. 4, we can find the
band gaps numbers of $TE$ wave decrease, and $TM$ wave band gaps
numbers decrease and become narrow with the function coefficient
$k$ decreasing. Comparing with the conventional photonic crystals
FIG. 1, we can find the band gaps numbers of $TE$ wave decrease,
and become wider, and the $TM$ wave band gaps numbers decrease and
become narrow. Comparing with FIG. 4 and FIG. 2, FIG. 5 and FIG.
3, we can find the band gaps of function coefficient $k<0$ are
wider than $k>0$. So, we can obtain the appropriate band gaps
structures by selecting the different function from of dielectric
constant.

\vskip 10pt {\bf 5. Conclusion} \vskip 10pt

In this paper, we have firstly proposed two-dimensional function
photonic crystals, and calculate the band gaps structures of $TE$
and $TM$ wave, and find the band gaps of two-dimensional function
photonic crystals are different from the two-dimensional
conventional photonic crystals. when the functions form of
dielectric constants and the crystal structure are different, the
band gaps structure should be changed, which can be designed into
the appropriate band gaps structures by the two-dimensional
function photonic crystals.

\vskip 12pt {\bf 6.  Acknowledgment} \vskip 12pt

This work was supported by the National Natural Science Foundation
of China (no.61275047), the Research Project of Chinese Ministry
of Education (no.213009A) and Scientific and Technological
Development Foundation of Jilin Province (no.20130101031JC).

\end{document}